\documentclass{ieeeaccess}

\usepackage[utf8]{inputenc}
\usepackage[T1]{fontenc}
\usepackage{textcomp}
\usepackage{xcolor}
\usepackage{amsmath,amssymb,amsfonts}
\usepackage{graphicx}
\usepackage{subcaption}
\usepackage[numbers,sort&compress]{natbib}

\def\BibTeX{{\rm B\kern-.05em{\sc i\kern-.025em b}\kern-.08em
    T\kern-.1667em\lower.7ex\hbox{E}\kern-.125emX}}

\graphicspath{{figures/}}

\begin{document}

\history{Date of publication xxxx 00, 0000, date of current version xxxx 00, 0000.}
\doi{10.1109/TQE.2020.DOI}

\title{Adaptive Resource and Memory Control for Stability in Quantum Entanglement Distribution}

\author{
\uppercase{Nicol\`o Lo Piparo}\authorrefmark{1},
\uppercase{William J. Munro}\authorrefmark{1},
and \uppercase{Kae Nemoto}\authorrefmark{1},\authorrefmark{2}
}

\address[1]{Okinawa Institute of Science and Technology Graduate University, Onna-son, Okinawa 904-0495, Japan}
\address[2]{National Institute of Informatics, 2-1-2 Hitotsubashi, Chiyoda-ku, Tokyo 101-8430, Japan}


\markboth
{Lo Piparo \headeretal: Adaptive Resource and Memory Control for Stability in Quantum Entanglement Distribution}
{Lo Piparo \headeretal: Adaptive Resource and Memory Control for Stability in Quantum Entanglement Distribution}

\corresp{Corresponding author: Nicol\`o Lo Piparo (email: nicopale@gmail.com)}

\begin{abstract}
We investigate congestion-aware control of quantum repeater nodes
operating under stochastic traffic and finite memory coherence. Entanglement
generation is modeled as a probabilistic process producing Werner
states subject to depolarizing memory decoherence, while entanglement
requests arrive according to Poisson and bursty ON--OFF processes.
Using a queueing-theoretic framework, we couple physical-layer memory
dynamics with congestion-dependent service behavior to analyze stability,
delay, and fidelity trade-offs. Operating regimes are characterized
in terms of the load parameter, showing that fixed cutoff policies
impose a fundamental fidelity--latency trade-off together with strict
stability limits. Queue-aware adaptive control strategies are then
introduced that dynamically adjust memory cutoff times and the number
of parallel entanglement-generation channels. Cutoff adaptation restores
stability near critical load by trading fidelity for service capacity,
whereas resource scaling increases capacity without degrading entanglement
quality. Under bursty traffic, joint adaptation suppresses delay spikes
while activating additional channels only during congestion periods.
The framework is further extended to a two-user shared-resource scenario
in which independent traffic flows compete for a common resource pool.
Stability is determined by aggregate load, while adaptive resource
redistribution stabilizes queues that diverge under fixed partitioning.
These results provide a queue-aware congestion-control perspective
for adaptive resource management in quantum networks.
\end{abstract}


\titlepgskip=-15pt
\maketitle

\section{Introduction}

Quantum networks rely on the distribution of entangled states between
 distant nodes to enable applications such as quantum key distribution
\citep{QKD1,QKD2,QKD3,QKD4}, distributed sensing \citep{DS1,DS2,DS3,DS4},
quantum repeater architectures \citep{Q_REPEATER_arch1,Q_REPEATER_arch2,Q_REPEATER_arch3,Q_REPEATER_arch4},
and large-scale quantum networking proposals \citep{Q_internet1,Q_internet2,Q_internet3,Q_internet4}.
In many practical implementations, entanglement generation is probabilistic
\citep{ENT_GEN,ENT_GEN3,ENT_GEN4} and entangled states must be temporarily
stored in quantum memories before being consumed \citep{Qmemories1,Qmemories2}.
During storage, decoherence degrades the quality of the stored states
\citep{Qmemories3,Qmemories4,Qmemories5,Qmemories6,Qmemories7}, creating
a fundamental trade-off between entanglement fidelity, latency, and
throughput \citep{cutOFF1,cutOFF2,RATEVSF_toff1,THROUGH_latency}.

A common approach to mitigate memory decoherence is to impose a finite
cutoff time: entangled states that have been stored longer than a
prescribed time are discarded rather than used. Such cutoff strategies
are known to improve the average fidelity of delivered entanglement,
but they also reduce the effective service rate of the system and
may increase waiting times. As a result, the choice of the cutoff
time plays a central role in determining overall system performance
\citep{cutOFF1,cutOFF2,cutOFF3,cutOFF4,cutOFF5}.

Cutoff policies have been studied extensively in repeater models as
a physical-layer mechanism for mitigating memory decoherence and optimizing
end-to-end rate or fidelity under prescribed operating conditions
\citep{cutOFF1,cutOFF6,cutOFF7,Q_REPEATER_arch5}. In most such studies,
the cutoff is treated as a static protocol parameter chosen offline
for a fixed architecture and traffic condition. By contrast, we study
cutoff control as a congestion-aware systems mechanism in the presence
of stochastic demand. The central question here is therefore not which
fixed cutoff optimizes a repeater in isolation, but how cutoff and
generation resources interact with queue buildup, latency, and stability
when entanglement requests arrive randomly and may be bursty \citep{ENT_GEN2,ENT_GEN5,BURSTY1,BURSTY2}. Existing
work on entanglement request scheduling and resource allocation in
quantum networks typically assumes fixed service characteristics.
In contrast, here the service process itself depends on physical-layer
mechanisms---namely memory decoherence and cutoff policies---and
therefore directly determines the effective service capacity. A key
step in our approach is to represent the repeater node as a service
system with a random service time $T$, which captures in a coarse
but useful way the combined effects of probabilistic entanglement
generation, memory decoherence, and cutoff-induced restarts. This
representation enables a direct characterization of system behavior
through the load parameter $\rho$, linking physical-layer parameters
to queue stability. From this perspective, the cutoff time is not
only a fidelity-control parameter but also a congestion-control mechanism:
by adjusting the cutoff parameter, the effective service capacity
is dynamically regulated in response to stochastic demand, while the
number of parallel generation channels provides an additional capacity-scaling
dimension.

In this work, we study entanglement distribution from a queueing-theoretic
perspective, explicitly accounting for stochastic arrivals, finite
service capacity, and memory decoherence. We model entanglement requests
as Poisson processes and describe the system as a single-server queue
with controllable service parameters. Entangled states are generated
as Werner states with finite visibility and undergo depolarizing decoherence
while stored in memory. The quality of the delivered states is quantified
by their fidelity, while performance is evaluated in terms of queue
length and delay.

We first analyze the behavior of the system under fixed control parameters,
focusing on three operating regimes determined by the load parameter
$\rho$: subcritical, critical, and supercritical. In the subcritical
regime, a finite cutoff time improves fidelity at the cost of increased
delay. In the critical regime, cutoff control can stabilize the queue,
but only by accepting a reduction in fidelity. In the supercritical
regime, cutoff control alone cannot prevent divergence of the queue,
although it reduces the growth rate of delay and backlog.

We then consider adaptive control strategies \citep{ADAPTIVEcontrol1,ADAPTIVEcontrol2,ADAPTIVEcontrol3},
in which system parameters are adjusted dynamically based on the observed
queue state. Two forms of adaptation are investigated: varying the
cutoff time and varying the number of parallel entanglement-generation
resources. We show that adapting the cutoff time alone has limited
effectiveness under high load, whereas adapting the number of resources
can stabilize the system in the critical regime without degrading
fidelity, at the expense of increased resource usage.

Finally, we extend the analysis to bursty arrival processes, modeled
by ON--OFF modulated Poisson traffic. In this setting, even systems
that are subcritical on average may experience transient overloads.
We demonstrate that adaptive strategies significantly reduce delay
spikes and fidelity degradation during bursts, compared to fixed-parameter
policies.

In addition to single-user operation, we extend the analysis to a
minimal multi-user configuration in which two independent traffic
flows share a common pool of entanglement-generation resources. This
setting captures a fundamental architectural feature of quantum networks:
physical resources at a repeater node must be dynamically partitioned
\citep{Q_routing1,Q_routing2,Q_routing3,Q_routing4} among competing
requests. We show that the total offered load determines global stability,
while traffic imbalance induces user-dependent fidelity and delay
behavior under adaptive control.

From a systems perspective, the goal of this work
is not to optimize a fixed repeater protocol or design optimal controllers,
but to understand how physical-layer parameters interact with queue
dynamics under stochastic demand. In particular, we aim to clarify
the structural role of cutoff policies and resource adaptation in
realistic operating conditions. Our results provide insight into when
cutoff strategies are beneficial, when they are insufficient, and
how congestion-aware resource redistribution can stabilize shared
quantum repeater nodes.

The main contributions of this work are:

1. Queue-coupled repeater service model. We introduce
a queueing-theoretic abstraction that couples quantum-memory decoherence
and cutoff policies with stochastic entanglement-request arrivals.
The resulting service model links physical-layer parameters (memory
lifetime, cutoff time, and generation resources) directly to queue
stability, latency, and entanglement fidelity.

2. Congestion-aware interpretation of cutoff control.
We reinterpret memory cutoff time as a dynamic congestion-control
mechanism rather than a static protocol parameter. This perspective
reveals how cutoff policies regulate the effective service rate and
therefore the stability condition of entanglement distribution under
stochastic demand.

3. Comparative analysis of adaptive cutoff and resource
scaling. We analyze two classes of queue-aware control strategies---adaptive
cutoff and adaptive channel allocation---and show that they play
complementary roles: cutoff adaptation trades fidelity for stability
near critical load, whereas resource scaling expands service capacity
without degrading entanglement quality.

4. Bursty-traffic and multi-user extensions. We
extend the framework to bursty ON--OFF traffic and to a shared-resource
two-user scenario. The results demonstrate how queue-aware adaptation
suppresses delay spikes under bursty demand and dynamically redistributes
generation resources among competing flows to maintain bounded operation
whenever aggregate load permits.

The remainder of the paper is organized as follows. Section II introduces
the physical and queueing model underlying the analysis. Section III
examines system behavior under fixed control policies and identifies
the fundamental operating regimes. Section IV develops queue-aware
adaptive control strategies and evaluates their performance under
stochastic traffic. Section V extends the framework to a shared-resource
multi-user scenario. Finally, Section VI concludes the paper and discusses
future directions.

\section{Physical and Queueing model}

\begin{figure}
\centering
\includegraphics[width=1\linewidth]{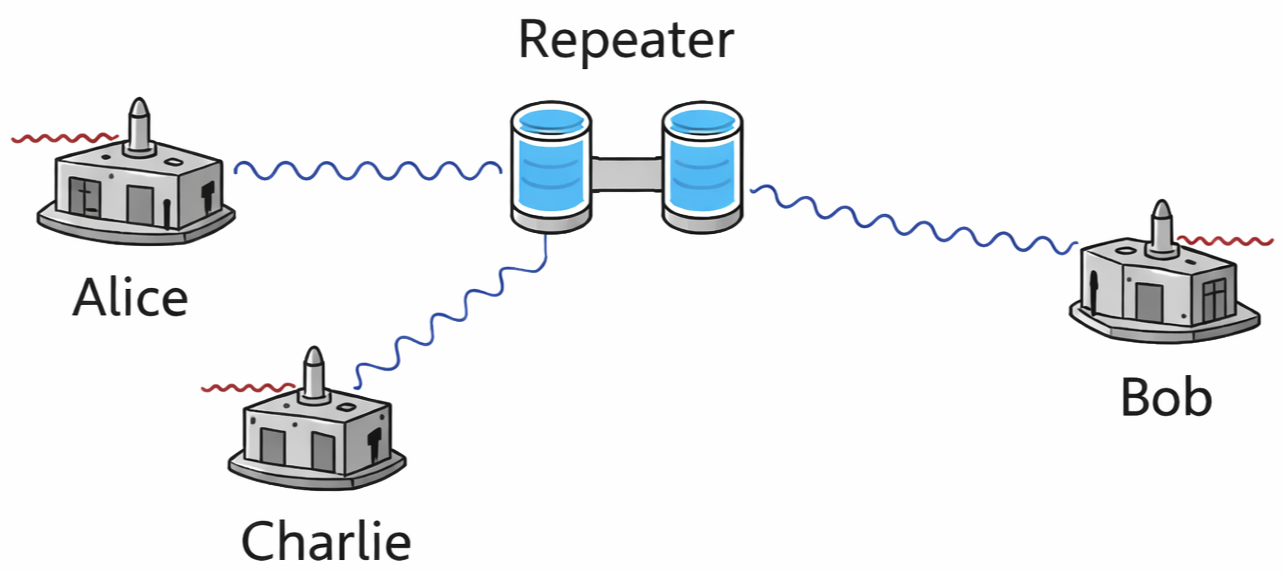}
\caption{\label{fig:Minimal-quantum-repeater}Minimal quantum repeater connecting Alice and Charlie to Bob separated by a distance $2L$.}
\end{figure}

We consider a minimal quantum-repeater architecture consisting of
two elementary entanglement-distribution links connected by an intermediate
node, where long-distance entanglement is established via entanglement
swapping, as shown in Fig.~\ref{fig:Minimal-quantum-repeater}. In
the baseline single-user scenario considered in this section, we focus
on a single stream of entanglement requests between the two end nodes
(Alice and Bob). The extension to multiple users sharing the same
repeater resources will be introduced in a later section. We assume
that entanglement swapping succeeds with probability one and introduces
no additional noise or delay. This allows us to focus exclusively
on the interplay between memory decoherence, cutoff time, and queueing
effects. Entanglement requests arrive randomly and are served by probabilistic
entanglement generation followed by storage and eventual delivery.
The model explicitly captures memory decoherence, finite service capacity,
and stochastic traffic. The assumptions of ideal swapping and homogeneous
elementary links are introduced to isolate the control and queueing
mechanisms governing stability and performance, rather than to provide
a hardware-complete repeater model. In more realistic architectures,
imperfect swapping, asymmetric link losses, or heterogeneous memory
parameters would modify the quantitative form of the service-time
function $\mathbb{E}[T(N,\tau)]$ and therefore shift the precise stability
boundaries. Here $T(N,\tau)$ denotes the random time required to
complete one end-to-end entanglement delivery under parameters $(N,\tau)$,
and $\mathbb{E}[T(N,\tau)]$ is its mean. However, the qualitative
behaviors identified in this work---including load regulation via
cutoff adaptation and congestion-driven resource scaling---depend
primarily on how control actions modify the effective service rate.
These structural trade-offs between fidelity, delay, and stability
are therefore expected to remain valid under more detailed physical
models, even though the exact operating thresholds will be hardware-dependent.

\subsection{Entanglement generation and memory decoherence}

Each successful entanglement generation attempt produces a two-qubit
Werner state of the form $\rho_{W}=v_{0}\,\lvert\Phi^{+}\rangle\!\langle\Phi^{+}\rvert+(1-v_{0})\,\frac{\mathbb{I}}{4}$,
where $v_{0}\in[0,1]$ denotes the initial visibility and $\lvert\Phi^{+}\rangle$
is a maximally entangled Bell state. The visibility $v_{0}$ accounts
for imperfections in state generation and transmission.

After generation, entangled states are stored in quantum memory while
waiting to be consumed. Memory decoherence is modeled as a depolarizing
process with characteristic coherence time $T_{2}$. If a state is
stored for a time $\tau$, its visibility decays as $v(\tau)=v_{0}\,e^{-\tau/T_{2}}$.
The fidelity of the delivered state with respect to the target Bell
state is then $F(\tau)=\frac{1}{4}+\frac{3}{4}v(\tau)^{2}$.

\subsection{Cutoff-time policy}

To limit decoherence, we introduce a cutoff time $\tau$: any entangled
state that has not been consumed within time $\tau$ after its generation
is discarded and not used for service. This cutoff improves the average
fidelity of delivered entangled states, but it reduces the effective
service capacity, since some generated states are wasted.

The cutoff time $\tau$ may be fixed or dynamically adjusted depending
on the control policy considered.

\subsection{Service model and effective rate}

Entanglement generation attempts occur at rate $R$, with each attempt
succeeding independently with probability $p_{\text{gen}}=e^{-L/L_{att}}$,
where $L$ is the distance of one elementary link and $L_{att}$ is
the attenuation length in standard optical fibers. We consider $N$
parallel, identical generation resources operating independently.
The effective generation rate is therefore $\mu(N)=NRp_{\text{gen}}$.
Throughout this work, unless otherwise stated, we fix the physical
parameters to representative values corresponding to a metropolitan-scale
elementary link. In particular, we consider an inter-node distance
$L=100\ \mathrm{km}$, an attempt rate $R=10^{6}\ \mathrm{s}^{-1}$,
a memory coherence time $T_{2}=10^{-5}\ \mathrm{s}$, and an initial
visibility $v_{0}=0.86$.

Given a cutoff time $\tau$, the effective service rate $S(N,\tau)$
accounts for both probabilistic generation and the probability that
a generated state survives until consumption. In the simulations,
this rate is computed via an analytically derived mean service time
$T_{\text{serv}}(N,\tau)\equiv\mathbb{E}[T(N,\tau)]$, with $S(N,\tau)=1/\mathbb{E}[T(N,\tau)]$.
The expression of the mean service time has been derived in Appendix
\ref{sec:Derivation-of-the}.

The impact of the cutoff time on system performance is mediated through
its effect on the mean service time. This dependence can be understood
heuristically as follows. Generation of a long-distance entangled
pair requires that two elementary links succeed within a finite coincidence
window. If one link succeeds first, its state is stored while waiting
for the second link. If the second link does not succeed within time
$\tau$, the stored state is discarded and the generation process restarts.

The service time therefore consists of repeated cycles of two-link
generation attempts. Reducing $\tau$ shrinks the coincidence window
and increases the probability that a partially generated pair is discarded
before completion. As a consequence, the expected number of restart
cycles grows, leading to an increase in the mean service time.

In the limit of large $\tau$, restarts become rare and the service
time approaches that of unconstrained parallel generation. In contrast,
for small $\tau$ the probability of successful coincidence within
a single cycle decreases, and the mean service time increases accordingly.
This restart mechanism underlies the trade-off between fidelity and
stability explored in the following sections.

\subsection{Service abstraction}

Entanglement generation involves parallel probabilistic attempts,
temporary storage of successful links in quantum memories, and possible
restart cycles induced by the cutoff policy. From the perspective
of the request queue, however, these microscopic processes are not
directly observable. What matters is only the time required to successfully
generate and deliver one end-to-end entangled pair.

We therefore describe the repeater as a service system whose service
time is the random variable $T(N,\tau)$, representing the total time
required to complete one entanglement-delivery event under parameters
$N$ (number of parallel generation channels) and $\tau$ (memory cutoff
time). Each successful entanglement delivery corresponds to a single
service completion event.

\subsection{Queueing model}

Entanglement requests arrive according to a stochastic arrival process.
In the baseline scenario, arrivals follow a Poisson process with rate
$\lambda$. Under this assumption, the system may be represented as
a single-server queue whose service time corresponds to the time required
to generate and deliver one end-to-end entangled pair. As discussed
above, the microscopic repeater dynamics consist of repeated probabilistic
attempts to generate elementary links, temporary storage of successful
links in quantum memories, and possible restart cycles due to the
cutoff policy. From the perspective of the request queue, however,
these internal processes are not directly observable. What matters
for queue evolution is only the time required to complete a single
entanglement-delivery event. We therefore describe the repeater as
a service system whose service time is the random variable $T(N,\tau)$,
representing the total time required to successfully generate and
deliver one entangled pair under parameters $N$ and $\tau$. Each completed
entanglement delivery corresponds to a service completion event.

Under Poisson arrivals, the service abstraction induced by the physical
generation process is naturally a renewal queue with independent and
identically distributed service times $T(N,\tau)$, and may therefore
be viewed as an M/G/1 model at the level of completed end-to-end pair
deliveries. In the numerical experiments below, however, we do not
simulate the full renewal service process directly. Instead, we employ
a rate-based approximation in which the instantaneous service capacity
is represented by the mean service rate
\[
S(N,\tau)=\frac{1}{\mathbb{E}[T(N,\tau)]}.
\]

Under this approximation, queue evolution is described by a discrete-time
update rule in which arrivals occur according to the stochastic arrival
process while service capacity is determined by the mean rate $S(N,\tau)$.

This approximation preserves the dependence of service capacity on
the physical control parameters $N$ and $\tau$, and is sufficient
for identifying operating regimes, stability transitions, and the
qualitative behavior of the adaptive control policies considered in
this work. This corresponds to a fluid mean-field approximation of
the underlying M/G/1 queue, which captures stability transitions and
control behavior while neglecting service-time variability effects.

Because the full distribution of $T(N,\tau)$ is not explicitly simulated,
the model does not capture higher-order waiting-time effects associated
with the variance of the service time. Consequently, quantitative
delay values near the critical load should be interpreted as approximate
indicators of regime behavior rather than exact M/G/1 waiting-time
predictions. Within this abstraction, the stability condition of the
queue remains governed by the standard load parameter $\rho=\lambda\mathbb{E}[T(N,\tau)]$,
with stable operation occurring when $\rho<1$. The adaptive policies
studied in the following sections regulate the system by modifying
$N$ and $\tau$, thereby altering the effective service rate and shifting
the operating point relative to this stability boundary.

\subsection{Traffic regimes}

System behavior is largely determined by the load parameter $\rho=\lambda\,\mathbb{E}[T(N,\tau)]$.

We distinguish three operating regimes:

\begin{itemize}
\item Subcritical regime ($\rho<1$): the queue is stable.
\item Critical regime ($\rho\approx1$): the system lies at the stability boundary.
\item Supercritical regime ($\rho>1$): the queue diverges over time.
\end{itemize}

These regimes are explored in the simulations to assess how cutoff
policies and resource adaptation affect stability, fidelity, and delay.

\subsection{Bursty arrival model}

To model traffic variability, we also consider bursty arrivals using
an ON--OFF modulated Poisson process. The arrival rate switches between
\[
\lambda(t)=\begin{cases}
\lambda_{\text{on}}, & \text{ON state},\\
\lambda_{\text{off}}, & \text{OFF state}.
\end{cases}
\]
with probabilistic transitions between states. This model captures
transient overloads even when the average arrival rate is subcritical.
Under this traffic model the arrival process becomes a Markov-modulated
Poisson process, corresponding to an MMPP/G/1 queue.

\subsection{Control policies}

We study and compare the following control strategies:

1. Fixed policy: both $N$ and $\tau$ are constant.

2. Adaptive cutoff: $\tau$ is adjusted dynamically based on the queue
state, while $N$ is fixed.

3. Adaptive resources: $N$ is adjusted dynamically, while $\tau$
is fixed.

4. Joint adaptation: both $N$ and $\tau$ are adjusted.

Performance is evaluated in terms of queue length, delay, fidelity,
and resource usage.

These feedback policies regulate the effective load of the queue by
dynamically adjusting the parameters that determine the service time.
When the queue grows, indicating that the arrival rate approaches
the service capacity, the controller increases either the cutoff time
$\tau$ or the number of parallel channels $N$, thereby increasing
the effective service rate $S(N,\tau)=1/\mathbb{E}[T(N,\tau)]$ and
pushing the load parameter $\rho=\lambda\mathbb{E}[T(N,\tau)]$ back
toward the stable regime. Conversely, when the queue becomes small,
the controller relaxes these parameters, allowing operation at higher-fidelity
or lower-resource settings. The detailed adaptive control framework
is described in Section IV.

The analysis above relies on several simplifying assumptions, including
ideal entanglement swapping, identical elementary links, and negligible
classical signaling delays. These assumptions allow the service time
$T(N,\tau)$ to be expressed in a compact form and help isolate the
impact of adaptive cutoff control on queue stability and delay. In
more realistic repeater architectures, additional effects such as
imperfect swapping operations, classical communication latency, or
heterogeneous link generation rates would modify the statistics of
the service time. In the queueing abstraction, however, these effects
primarily alter the quantitative value and distribution of $T(N,\tau)$
without changing the qualitative structure of the model. In particular,
the system would still be described by an M/G/1 queue with a modified
service-time distribution, so the stability condition $\rho=\lambda\mathbb{E}[T]<1$
and the congestion mechanisms analyzed here remain unchanged.

\section{Fixed Policies and Operating Regimes}

Building on the physical and queueing framework developed in Section
II, we now analyze the system under fixed control policies, where
the number of parallel channels $N$ and the cutoff time $\tau$ are
held constant. This baseline analysis establishes the fundamental
operating regimes and stability boundaries that adaptive strategies
will later regulate dynamically.

\subsection{Fixed cutoff time}

\begin{figure}
\begin{centering}
\includegraphics[scale=0.6]{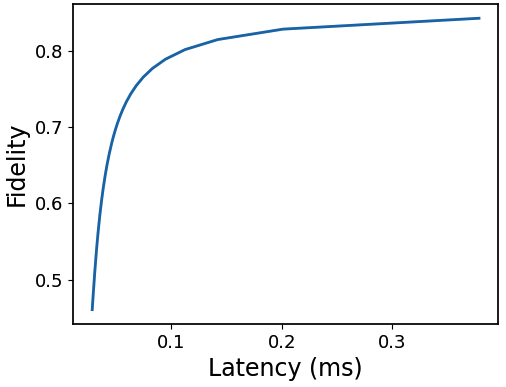}
\par\end{centering}
\caption{\label{fig:Steady-state-fidelity-versus}Steady-state fidelity versus
average latency for a fixed number of channels $N=5$, obtained by
varying the memory cutoff time $\tau$. }

\end{figure}

We first consider a fixed number of parallel channels $N$ and vary
the memory cutoff time $\tau$. Introducing a finite cutoff improves
steady-state fidelity by limiting memory dwell times, but increases
the mean service time through more frequent restart events. The resulting
fidelity--latency trade-off is shown in Fig. \ref{fig:Steady-state-fidelity-versus}.

\subsection{Operating Regimes Based on Traffic Load}

\begin{figure}
\begin{centering}
\includegraphics[scale=0.5]{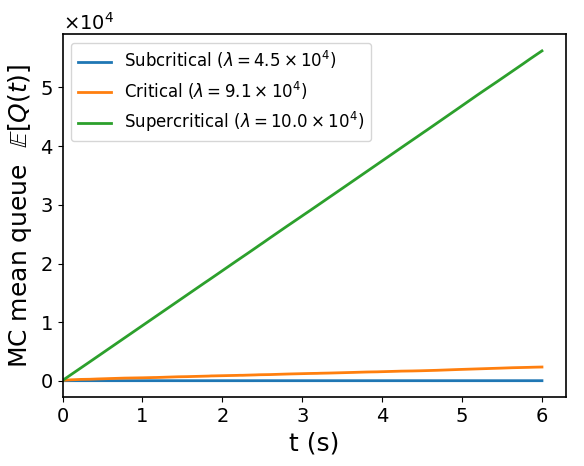}
\par\end{centering}
\caption{\label{fig:fixed_3 regimes}Monte Carlo mean queue length $\mathbb{E}[Q(t)]$
under the fixed policy for three arrival rates $\lambda\text{ with}N_{\text{fix}}=16$
and $\tau=0.5T_{2}$. For subcritical load ($\lambda=4.5\times10^{4}$),
the queue remains bounded and stabilizes. Near the critical regime
($\lambda=9.1\times10^{4}$), the queue exhibits slow growth, indicating
marginal stability. In the supercritical regime ($\lambda=10^{5}$),
the mean queue grows approximately linearly in time, signaling instability
due to arrival rate exceeding service capacity.}
\end{figure}
The operating regime is determined
by the load parameter $\rho=\lambda\,\mathbb{E}[T(N,\tau)]$. When
$\rho$< 1, the queue admits a stationary distribution and
the system is asymptotically stable. At $\rho$ = 1, the system
lies on the stability boundary: no stationary distribution exists
and the queue exhibits marginal drift. For $\rho>1$, the queue has
positive drift and diverges linearly in time. We distinguish three
qualitatively different operating regimes depending on the value of
$\rho$.

\textbf{Subcritical regime}
($\rho<1$)

In the subcritical regime, the system is stable
and the queue length remains bounded over time. This behavior is illustrated
in Fig. \ref{fig:fixed_3 regimes}, where the queue (blue curve) converges
to a finite steady value. In this regime, introducing a finite cutoff
time $\tau$ primarily acts as a fidelity-enhancement mechanism. By
discarding states that experience long storage times, memory decoherence
is limited and the steady-state fidelity of the distributed entangled
pairs increases. The corresponding increase in the mean service time
leads to a modest increase in the average delay, but does not affect
stability. The cutoff therefore primarily tunes the fidelity--latency
trade-off without affecting stability.

2. \textbf{Critical regime}
($\rho\approx1$)

At the critical point the system becomes highly
sensitive to the choice of the cutoff time. In Fig. \ref{fig:fixed_3 regimes}
(red curve) the Monte Carlo average $\mathbb{E}[Q(t)]$ grows slowly
in time and does not clearly settle to a steady value within the simulation
window. At $\rho\approx1$, the system lies at the stability boundary,
and no stationary queue distribution is expected in the asymptotic
sense. Over finite simulation windows, the drift may appear gradual
rather than strictly linear. The observed slow drift of $\mathbb{E}[Q(t)]$
therefore reflects operation in the vicinity of the critical load
rather than bounded steady-state behavior.

\textbf{3. Supercritical regime}
($\rho>1$)

In the supercritical regime the arrival rate exceeds
the effective service capacity determined by the fixed cutoff, as
shown by the red curve in Fig. \ref{fig:fixed_3 regimes}. The Monte
Carlo average $\mathbb{E}[Q(t)]$ grows approximately linearly in
time, indicating persistent overload. While the cutoff still limits
the memory dwell time and therefore bounds the fidelity, it cannot
restore stability.When $\lambda\mathbb{E}[T(N,\tau)]>1$, the queue
is unstable for that specific choice of $\tau$. However, for marginal
overload conditions, increasing the cutoff time may reduce the effective
load and restore stability. When the arrival rate exceeds the maximum
achievable service capacity, no cutoff-based policy can stabilize
the system.

\subsection{Fixed Number of Channels vs Fixed Cutoff}

We next contrast cutoff control with resource scaling
by fixing the cutoff time $\tau$ and varying the number of parallel
entanglement-generation channels $N$.

Increasing $N$ increases the effective service
rate $S(N,\tau)$. As a consequence, increasing $N$ directly improves
queue stability and reduces delay by expanding the service capacity
available to meet the request rate $\lambda$. In the critical regime,
this resource scaling can restore stability without requiring any
reduction in entanglement quality.

The effect of $N$ on fidelity is comparatively
weaker: the fidelity increases with $N$ for small $N$ (through the
increase in $\mu(N)$) but rapidly saturates once $\mu(N)\tau$ is
sufficiently large. By contrast, the cutoff time $\tau$ directly
regulates memory dwell time: decreasing $\tau$ improves fidelity
but increases the mean service time through more frequent restarts.

Resource scaling primarily expands service capacity
and shifts the stability boundary, whereas cutoff control regulates
entanglement quality at the expense of service efficiency. The analysis
above reveals that fixed policies impose a rigid trade-off between
fidelity, delay, and stability. In particular, near the critical load,
small parameter changes can shift the system from stable to unstable
operation. This sensitivity motivates the introduction of adaptive
strategies that dynamically adjust operating parameters in response
to congestion. We therefore turn next to queue-aware adaptive control.

\section{Adaptive control policies}

In this section we move beyond fixed operating points
and investigate adaptive control strategies in which the physical
parameters of the repeater node are adjusted dynamically in response
to the instantaneous congestion level. The purpose of this analysis
is not to identify optimal control laws, but rather to assess to what
extent simple, queue-aware adaptation can improve performance and
robustness under time-varying traffic conditions.

As shown in Section III, fixed choices of the cutoff
time $\tau$ and the number of parallel channels $N$ define distinct
operating points characterized by different trade-offs between fidelity,
latency, and stability. Adaptive control may be viewed as a mechanism
to dynamically navigate this trade-off space as traffic conditions
evolve.

Throughout this section, the controller is assumed
to observe only the instantaneous queue length $Q(t)$. No prediction
of future arrivals and no explicit knowledge of the traffic statistics
are required.

\subsection{Queue-aware adaptive control framework}

We consider adaptive policies in which one or both
of the controllable parameters, namely the memory cutoff time $\tau$
and the number of parallel entanglement-generation channels $N$,
are adjusted as functions of the current queue length $Q(t).$

Conceptually, an increase in $Q(t)$ signals congestion
and motivates a response aimed at increasing the effective service
capacity of the system, while a small queue indicates that higher-fidelity
operating points may be sustained. In the following, we examine three
classes of adaptive policies:

1. adaptive cutoff control with fixed $N$, 

2. adaptive resource scaling with fixed $\tau$, 

3. joint adaptation of $\tau$ and $N$.

We emphasize that these policies are heuristic and intentionally simple,
serving to illustrate fundamental performance limits rather than optimal
control strategies. The objective is not to design
a provably optimal controller, but to examine whether minimal queue-aware
feedback is sufficient to regulate operation near the stability boundary
identified in Section II.

\subsubsection{Feedback laws}

The controller updates the operating parameters
at discrete times $t_{k}=k\Delta t$, using only the current queue
length $Q(t_{k})$. In the adaptive-cutoff experiments, the memory
cutoff is chosen according to the linear queue-feedback rule: $\tau(t_{k})=\mathrm{clip}\!\left\{ \tau_{0}+\kappa_{\tau}\bigl(Q(t_{k})-Q_{{\rm target}}\bigr),\tau_{\min},\tau_{max}\right\} ,$
where $\tau_{0}$ is a nominal cutoff value, $Q_{{\rm target}}$ is
a desired queue operating point, and $\kappa_{\tau}>0$ controls responsiveness.
This proportional feedback law provides a simple negative-feedback
mechanism: when the queue exceeds its target value, the cutoff time
is increased, which raises the effective service rate $S(N,\tau)=1/\mathbb{E}[T(N,\tau)]$
and pushes the load parameter $\rho=\lambda\mathbb{E}[T(N,\tau)]$
downward toward the stability region. Conversely, when the queue is
small, the cutoff relaxes toward its nominal value, allowing higher-fidelity
operation. In this sense, the controller continuously adjusts the
operating point in response to congestion, without requiring explicit
knowledge of the arrival rate.

When adaptive resource scaling is enabled, the number
of parallel entanglement-generation channels is updated as : $N(t_{k})=\mathrm{clip}\!\Bigl(\mathrm{round}\bigl(N_{0}+\kappa_{N}(Q(t_{k})-Q_{{\rm target}})\bigr),\ N_{\min},N_{\max}\Bigr).$
Here again, proportional queue feedback increases capacity during
congestion and reduces resource usage during low-load periods. Although
the controller gains $\kappa_{\tau}\text{ and }\kappa_{\ensuremath{N}}$
affect transient dynamics and oscillation amplitude, the qualitative
behaviors reported below --- stabilization near critical load, congestion
mitigation under bursts, and resource redistribution in the multi-user
case --- are robust across a broad range of gain values. The feedback
laws should therefore be interpreted as representative queue-aware
mechanisms rather than finely tuned control solutions.

\subsection{Adaptive cutoff control with Poissonian arrivals}

We first consider adaptive cutoff control in which
the number of parallel channels $N=16$ is fixed, while the cutoff
time $\tau$ is adjusted dynamically as a function of the queue length.
The controller increases $\tau$ when congestion builds up and decreases
it when the queue relaxes, thereby modulating the effective mean service
time. 

For Poissonian arrivals with constant rate $\lambda$,
this adaptive strategy exhibits behavior consistent with the operating
regimes identified in Section III. In
the subcritical regime, adaptive control produces only minor fluctuations
without qualitative change; results are omitted for brevity.

\begin{figure*}
\begin{centering}
\begin{tabular}{|c|c|c|}
\hline 
\includegraphics[scale=0.166]{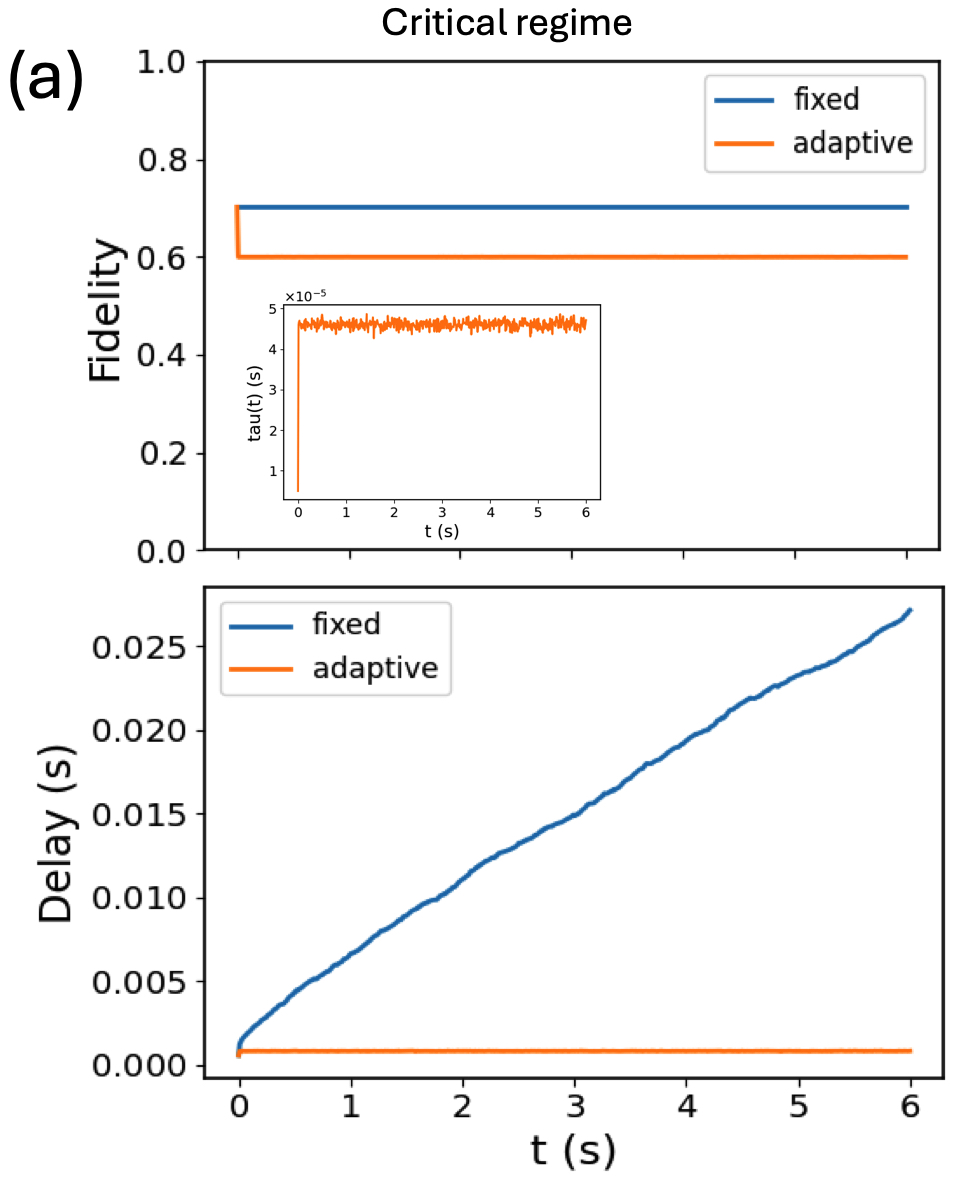} & \includegraphics[scale=0.168]{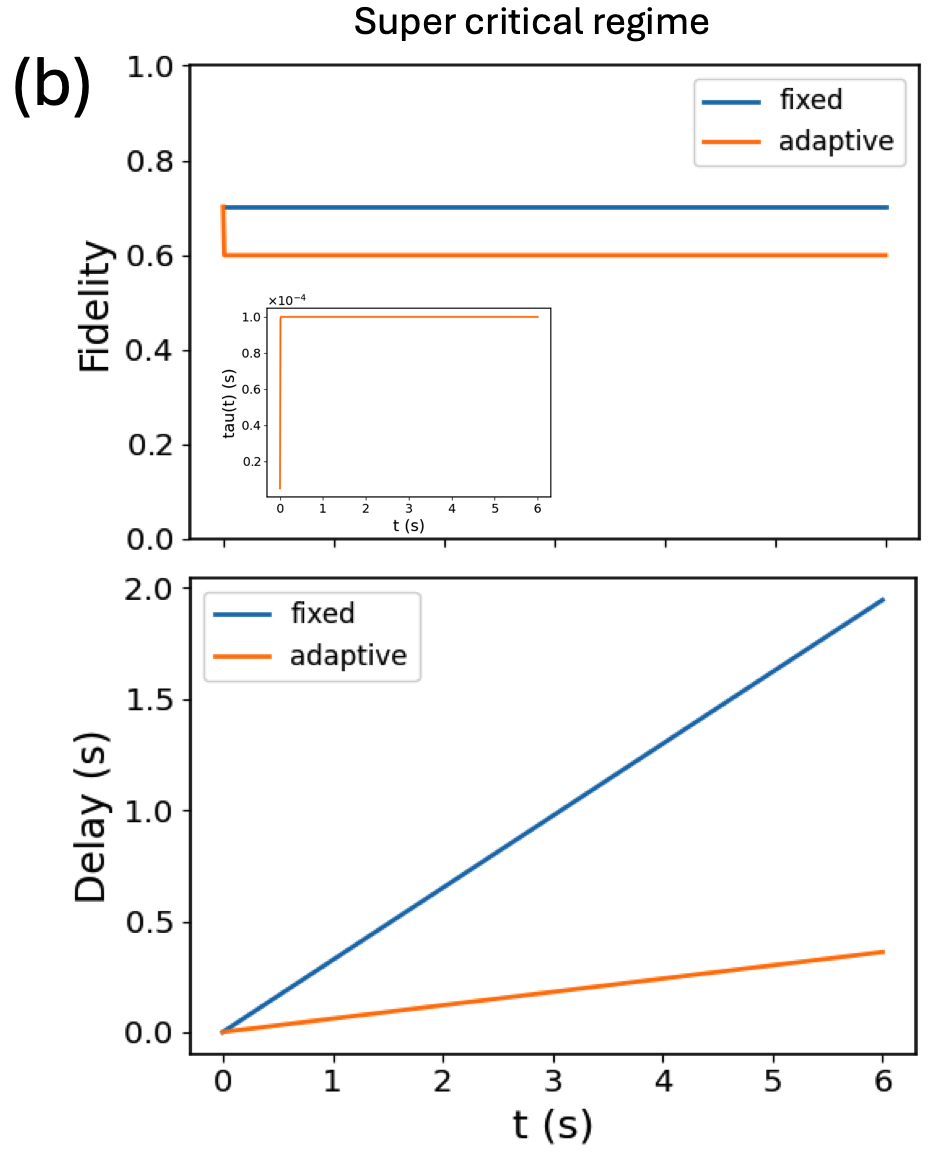} & \includegraphics[scale=0.168]{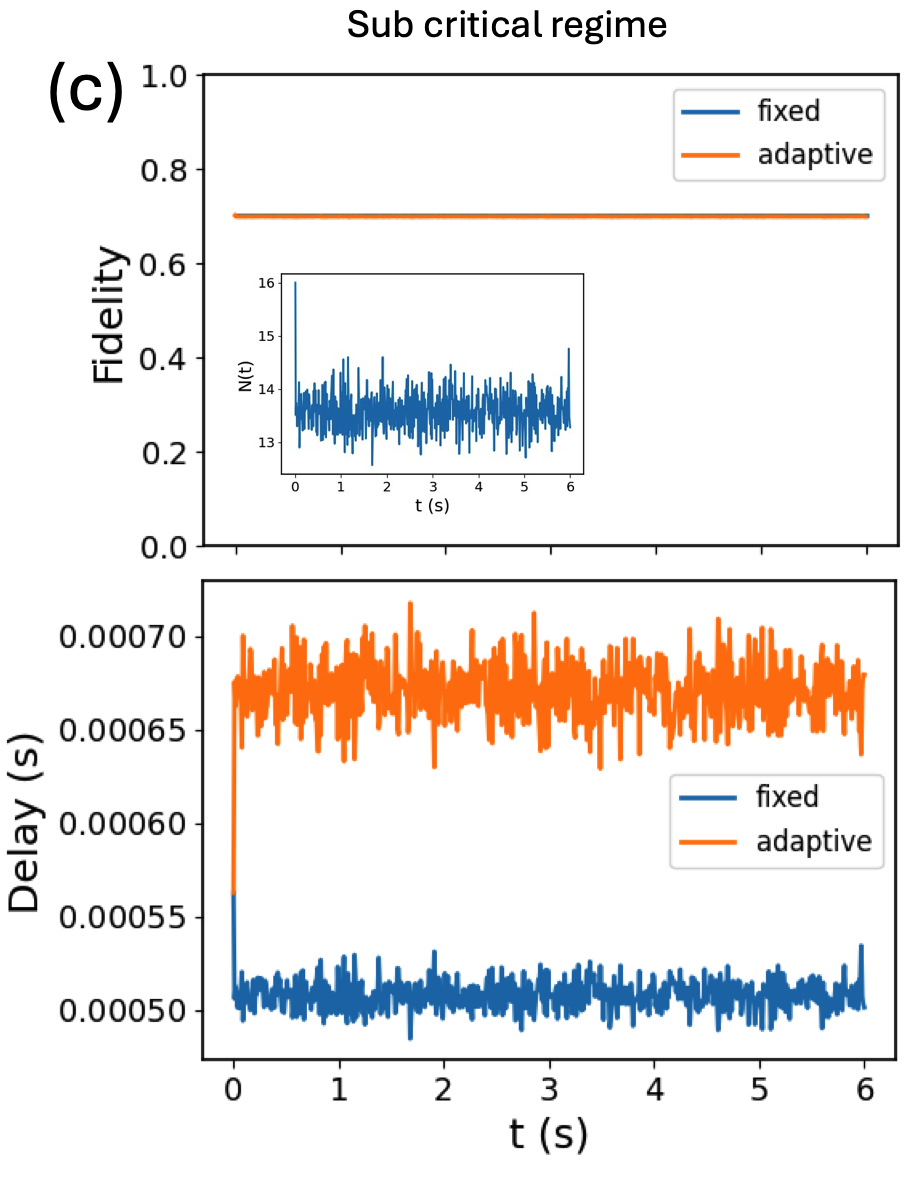}\tabularnewline
\hline 
\end{tabular}
\par\end{centering}
\caption{\label{fig:Comparison_critical}Comparison between a fixed-cutoff
baseline and a queue-aware adaptive policy across different traffic
regimes. Panels (a)--(c) report the temporal evolution of fidelity
(left) and delay (right) for the two policies.(a) Critical regime
with a fixed number of parallel attempts $N$. Inset (left): time
evolution of the adaptive cutoff $\tau(t)$. (b) Supercritical regime
with a fixed number of parallel attempts $N$. Inset (left): time
evolution of the adaptive cutoff $\tau(t)$. (c) Subcritical regime
with fixed cutoff $\tau$. Inset (left): time evolution of the adaptive
number of channels $N(t)$.}

\end{figure*}

In contrast, near the critical regime ($\lambda\mathbb{E}[T]\approx1$),
adaptive cutoff control can play a significant role, as show in Fig.
\ref{fig:Comparison_critical} (a). While a fixed cutoff may lead
to instability for certain parameter choices, dynamically increasing
$\tau$ in response to congestion can shift the system to operating
points with $\lambda\mathbb{E}[T(N,\tau)]<1$, thereby restoring bounded
delay. This stabilization mechanism comes at the cost of reduced fidelity
during congested periods, highlighting an explicit trade-off between
entanglement quality and queue stability. 

In the strongly supercritical regime, where the
arrival rate exceeds the maximum achievable service capacity even
in the limit of large $\tau$, adaptive cutoff control cannot restore
stability when $\lambda\mathbb{E}[T(N,\tau)]>1$ for all admissible
$\tau$. Nevertheless, increasing $\tau$ in response to congestion
can mitigate the growth rate of the queue and reduce delay relative
to fixed-cutoff operation, as shown in Fig. \ref{fig:Comparison_critical}
(b). This behavior reflects a fundamental capacity limitation rather
than a failure of the control policy. 

\subsection{Adaptive channel number with Poissonian arrivals}

We next consider adaptive resource scaling in which
the cutoff time $\tau$ is fixed, while the number of parallel generation
channels N is adjusted dynamically as a function of the queue length.
Arrivals follow a Poisson process and the system operates in the subcritical
regime, $\lambda\mathbb{E}[T]<1.$

Figure \ref{fig:Comparison_critical} (c) reports
the time evolution of fidelity, delay, and the channel allocation
$N(t)$. In this regime, both fixed and adaptive policies remain stable,
as expected. The fidelity traces are nearly indistinguishable, indicating
that modest fluctuations in $N(t)$ do not significantly affect entanglement
quality when the system is well below the stability boundary.

The delay under adaptive control is only marginally
higher than in the fixed case, reflecting small fluctuations in instantaneous
service capacity. The principal effect is reduced average channel
usage while maintaining comparable fidelity and bounded delay. In
subcritical conditions, resource adaptation therefore acts primarily
as an efficiency mechanism rather than a stability mechanism. In critical
or supercritical regimes, by contrast, the controller increases $N$
in response to congestion, whereas fixed operation eventually becomes
unstable.

\subsection{Sensitivity to feedback gains}

\begin{figure}
\begin{centering}
\includegraphics[scale=0.38]{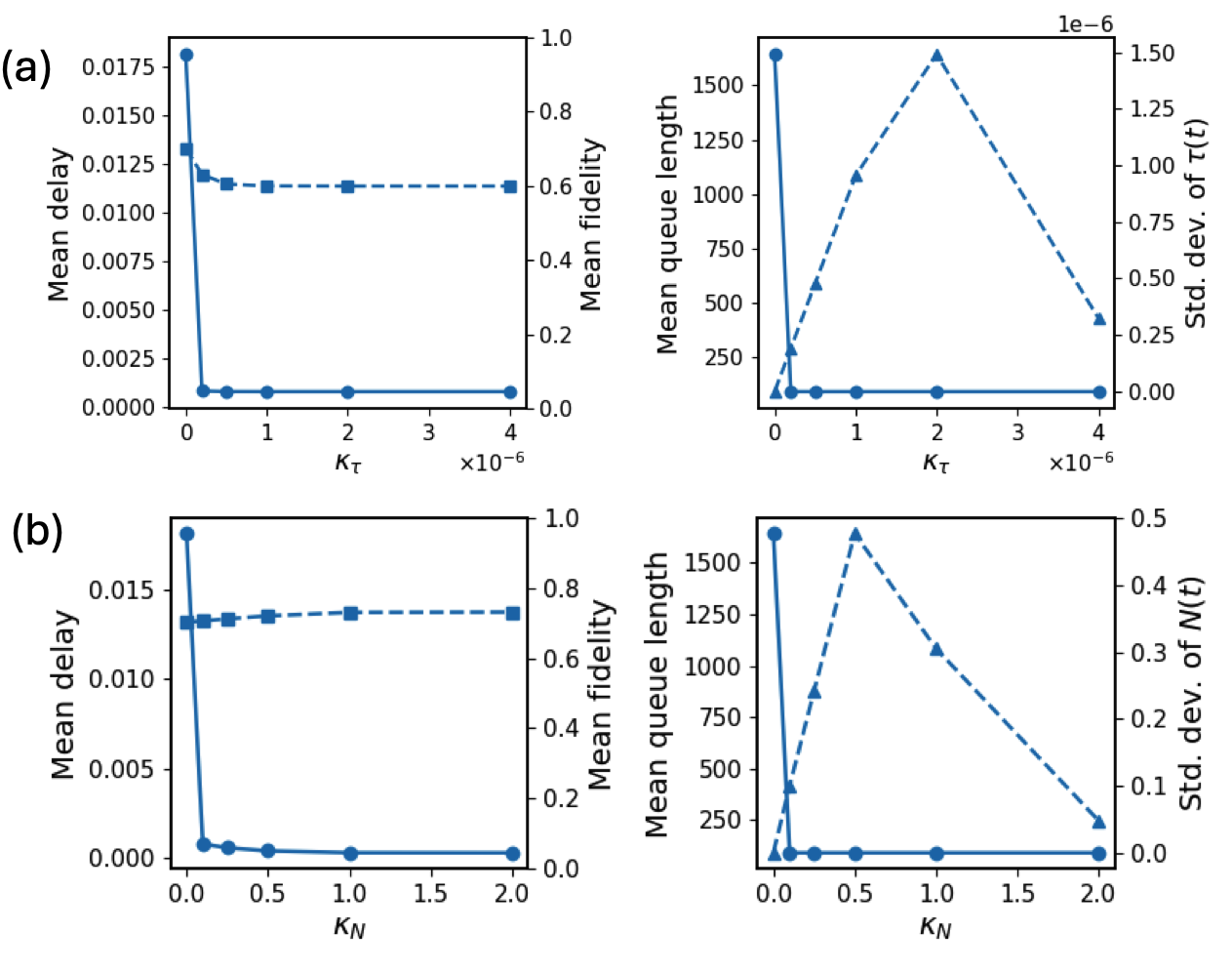}
\par\end{centering}
\caption{\label{fig:kappas_anal}(a) Cutoff control: time-averaged delay and
fidelity (left) and standard deviation of the cutoff signal $\tau(t)$
(right) as functions of the feedback gain $\kappa_{\tau}.$ (b) Resource
scaling: time-averaged delay and fidelity (left) and standard deviation
of the channel allocation $N(t)$ (right) as functions of the feedback
gain$\kappa_{N}$. }
\end{figure}

To assess whether the behavior of the adaptive controllers depends
sensitively on the proportional feedback gains, we performed parameter
sweeps over the cutoff-control gain$\kappa_{\tau}$ and the resource-scaling
gain $\kappa_{N}$ in a representative near-critical operating regime.
The resulting performance metrics and control-signal variability are
summarized in Fig. \ref{fig:kappas_anal}. For each gain value, we
measured time-averaged queue length, delay, and fidelity, together
with the variability of the corresponding control signal.

For cutoff control, the system without feedback ($\kappa_{\tau}=0$)
operates slightly above the stability boundary, resulting in substantial
queue buildup and large delays. Introducing even a small positive
gain rapidly stabilizes the queue, reducing delay by more than an
order of magnitude. Once stabilization is achieved, further increases
in $\kappa_{\tau}$ produce only minor changes in the average performance
metrics. At the same time, increasing the gain leads to larger variations
of the cutoff signal $\tau(t)$ over time, reflecting a more aggressive
response to instantaneous congestion. For sufficiently large gains,
however, this variability decreases again as the control signal increasingly
operates near its clipping bounds. In contrast to classical feedback
systems, no pronounced oscillatory behavior is observed at large gains.
This is due to the rate-based service abstraction and the presence
of saturation in the control law, which together yield a strongly
damped response in which large gains primarily lead to saturation
rather than sustained oscillations.

For resource scaling, the qualitative behavior with respect to stabilization
is similar: any positive gain$\kappa_{N}$ is sufficient to suppress
the queue growth observed in the uncontrolled case. However, the impact
on performance differs from cutoff control. Increasing $\kappa_{N}$
improves both delay and fidelity. By increasing the number of parallel
generation channels during periods of congestion, the controller raises
the effective service rate, thereby reducing queueing delay. At the
same time, faster generation reduces the storage time of elementary
links in quantum memories, leading to a slight increase in fidelity.
As in the cutoff case, once the system is stabilized, performance
becomes largely insensitive to the precise value of the gain. The
variability of the control signal $N(t)$ increases for intermediate
gains but decreases again for larger gains as saturation effects become
dominant.

Overall, these results demonstrate that the qualitative behavior of
the adaptive controllers does not rely on fine tuning of the feedback
gains. A broad intermediate range of gain values yields similar stabilization
and performance characteristics, indicating that the proposed feedback
laws should be interpreted as representative queue-aware mechanisms
rather than carefully optimized controllers.

\subsection{Adaptive cutoff control under bursty arrivals}

We now consider adaptive cutoff control in the presence
of time-varying traffic, modeled as bursty Poisson arrivals. Specifically,
the arrival process alternates between periods of low activity and
high-intensity bursts while preserving a prescribed long-term average
arrival rate $\lambda_{{\rm avg}}$. Such fluctuations are representative
of realistic network conditions and are not captured by stationary
Poisson models. 

The arrival process is modeled as a two-state ON--OFF
Poisson process. During OFF periods the arrival rate is $\lambda_{{\rm off}}=0$,
while during ON periods requests arrive according to a Poisson process
with rate $\lambda_{{\rm on}}$. Transitions from the ON state to
the OFF state occur with probability $p_{10}$, while transitions
from OFF to ON occur with probability $p_{01}$. The stationary ON-state
occupation probability is given by $p_{{\rm on}}=\frac{p_{01}}{p_{01}+p_{10}}$,
resulting in a long-term average arrival rate $\lambda_{{\rm avg}}=p_{{\rm on}}\lambda_{{\rm on}}$.

For moderate burst intensities with average arrival
rate near the critical regime, adaptive cutoff control produces the
same qualitative behavior observed in the stationary case: during
burst periods the controller temporarily increases the cutoff time
$\tau$, reducing delay at the cost of a transient decrease in entanglement
fidelity. Once the burst subsides, the cutoff returns to smaller values
and the system recovers its higher-fidelity operating point. Detailed
results for this mild burst regime are reported in Appendix B. 
\begin{figure}
\begin{centering}
\begin{tabular}{|c|c|}
\hline 
\includegraphics[scale=0.12]{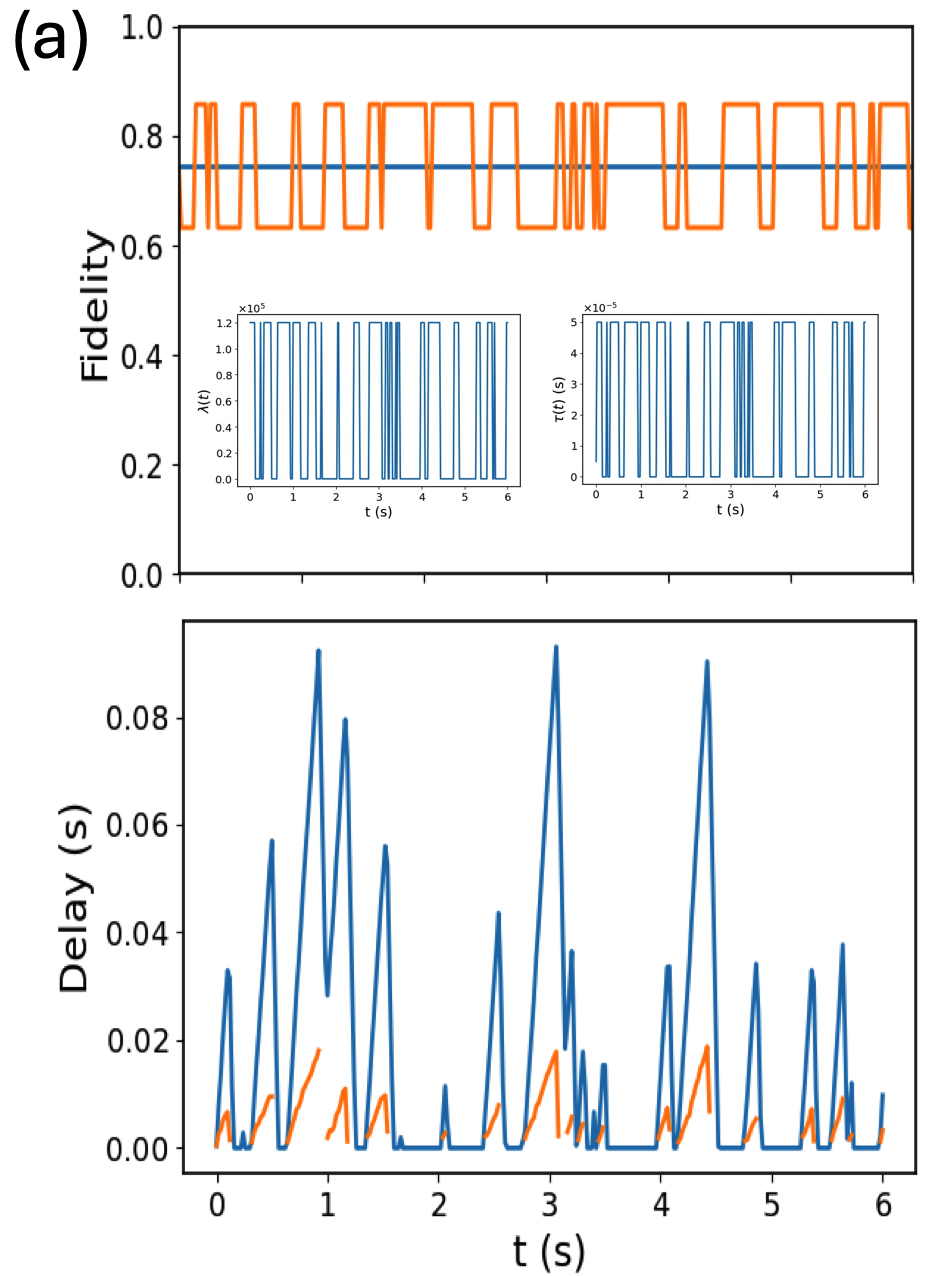} & \includegraphics[scale=0.12]{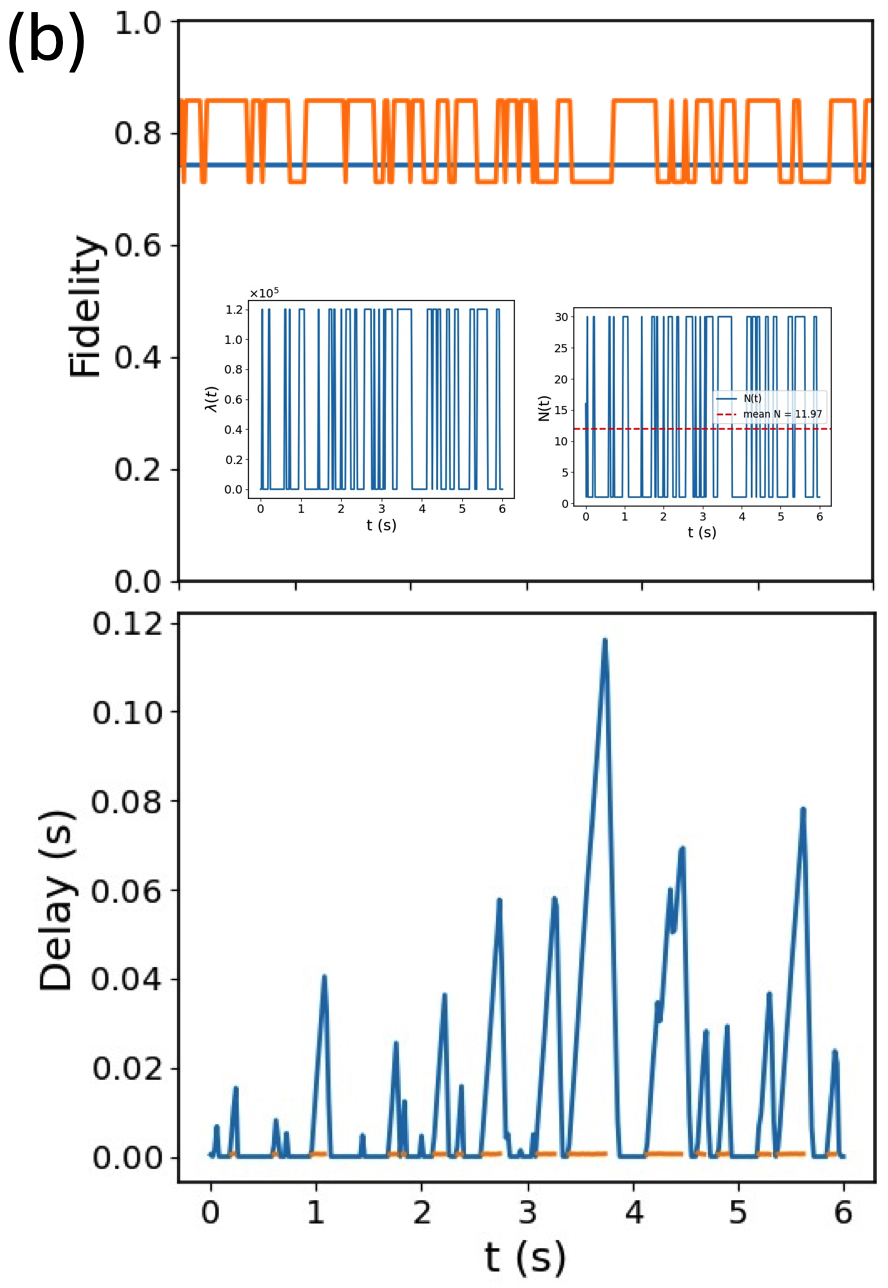}\tabularnewline
\hline 
\end{tabular}
\par\end{centering}
\caption{\label{fig:bursty2}Adaptive control under bursty Poisson arrivals
with supercritical ON periods. Panels (a)--(b) show the temporal
evolution of fidelity (left) and delay (right) for fixed operation
(blue) and adaptive strategies (red).(a) Adaptive cutoff control with
fixed number of parallel channels $N$. The left inset shows the instantaneous
arrival rate $\lambda(t)$, modeled as an ON--OFF Poisson process
with supercritical burst intensity$\lambda_{\mathrm{on}}=1.2\times10^{5}$
and $\lambda_{\mathrm{off}}=0$. The right inset shows the time evolution
of the cutoff $\tau(t)$. (b) Joint adaptive control of cutoff $\tau(t)$
and number of parallel channels $N(t)$. The left inset shows the
same arrival process$\lambda(t)$ as in panel (a). The right inset
shows the time evolution of the number of active channels $N(t)$.}

\end{figure}

We next consider supercritical ON periods with $\lambda_{ON}$
$\mathbb{E}[T]>1$. Figure \ref{fig:bursty2} illustrates this regime
for$\lambda_{{\rm on}}=1.2\times10^{5},p_{10}=0.01$, corresponding
to an average arrival rate $\lambda_{{\rm avg}}=4.5\times10^{4}$.
As shown in Fig.\ref{fig:bursty2} (a), extended bursts drive transient
overload. Adaptive cutoff again mitigates delay growth at the expense
of reduced fidelity, but cannot eliminate instability when instantaneous
load exceeds capacity.

As in the previous bursty example, adaptive cutoff
control does not eliminate the impact of supercritical traffic, but
it significantly attenuates delay peaks relative to fixed operation.
These results reinforce the interpretation of adaptive cutoff control
as a congestion-mitigation mechanism that dynamically trades entanglement
quality for improved temporal performance under highly variable traffic
conditions. 

Finally, we consider a joint adaptive strategy in
which both the cutoff time $\tau$ and the number of parallel entanglement-generation
channels $N$ are adjusted dynamically in response to congestion.
While this comparison is not \textquotedblleft fair\textquotedblright{}
in the sense that additional physical resources are being introduced,
it serves to clarify the complementary roles played by cutoff control
and resource scaling. Figure \ref{fig:bursty2} (b) illustrates this
behavior for the same bursty traffic configuration considered previously.
As shown in the left inset, the arrival process exhibits supercritical
bursts. The center inset shows that joint adaptation substantially
suppresses delay spikes relative to fixed operation, similarly to
cutoff-only adaptation. The key difference is visible in the main
fidelity trace: although the adaptive fidelity still decreases during
bursts, it never drops below the fixed-policy baseline.

This behavior is explained by the temporary increase
in the number of active channels $N(t)$, shown in the right inset.
During burst periods, increasing N raises the aggregate entanglement-generation
rate and compensates for the longer memory dwell times induced by
increasing $\tau$. As a result, fidelity degradation caused by cutoff
adaptation is mitigated without sacrificing queue stability. Importantly,
the increase in N is not permanent. During low-traffic periods the
controller reduces the number of active channels, allowing the system
to return to a resource-efficient operating point. This highlights
a key insight: maintaining high fidelity under bursty traffic generally
requires acting on both control knobs. 

An additional observation concerns the average resource
usage under joint adaptive control. In the supercritical burst scenario
of Fig. \ref{fig:bursty2} (b), the time-averaged number of active
channels over the full simulation window was approximately 12, lower
than the fixed baseline $N=16$. This indicates that joint adaptation
does not simply increase resources permanently, but rather activates
additional channels only during congestion intervals and releases
them during low-traffic periods. As a result, the system achieves
improved delay suppression and fidelity preservation while reducing
average resource consumption compared to static over-provisioning.
This highlights the potential of queue-aware control to improve not
only performance but also hardware efficiency. 

Having characterized adaptive behavior in the single-user
case, we now examine how congestion-aware control operates when multiple
traffic flows compete for shared physical resources.

\section{Two Users Sharing a Common Resource Pool}

The adaptive mechanisms studied thus far regulate
congestion for a single traffic flow. In practice, however, quantum
repeater nodes must serve multiple users simultaneously, with shared
physical resources. We therefore extend the framework to a minimal
two-user scenario in which independent traffic flows compete for a
common pool of entanglement-generation channels, as illustrated in
Fig. \ref{fig:Minimal-quantum-repeater}.

Specifically, two users, Alice and Charlie, request
end-to-end entanglement with Bob through the same repeater node. The
repeater--Bob segment provides a total of $N_{\mathrm{tot}}$ parallel
entanglement-generation channels that must be dynamically allocated
between the two flows. At any time t, the allocation satisfies $N_{A}(t)+N_{C}(t)=N_{\mathrm{tot}},\qquad N_{u}(t)\ge N_{\min},\;u\in\{A,C\}$,
> where $N_{A}(t)$ and $N_{C}(t)$ denote the number of channels
assigned to Alice--Bob and Charlie--Bob requests, respectively.

Traffic is modeled as two independent Poisson arrival
processes with rates $\lambda_{A}$ and $\lambda_{C}$. Each flow
maintains its own queue $Q_{A}(t)$ and $Q_{C}(t)$, while service
is coupled through the shared constraint on $N_{{\rm tot}}$. Unless
stated otherwise, we keep the same physical-layer parameters as in
the single-user case (e.g., L, R, $T_{2}$, and $v_{0}$), and we
continue to assume ideal swapping to isolate the queueing/control
effect of resource sharing. To classify the operating regime in the
two-user setting, it is convenient to introduce a total load parameter
$\rho_{{\rm tot}}\;=\;(\lambda_{A}+\lambda_{C})\,\mathbb{E}[T(N_{{\rm tot}},\tau)]$
, where $N_{{\rm tot}}$ is the total number of shared channels on
the repeater--Bob segment. The quantity $\mathbb{E}[T(N,\tau)]$
is computed from the same service-time model introduced in Secs. II--III.

This definition collapses the regime classification
to the aggregate arrival rate, independently of how traffic is split
between the two users. As in the single-user case: 

\textbullet{} $\rho_{{\rm tot}}<1$: subcritical
regime (stable),

\textbullet{} $\rho_{{\rm tot}}\approx1$: critical
regime, 

\textbullet{} $\rho_{{\rm tot}}>1$: supercritical
regime (unstable).

Importantly, stability is now determined by the
total offered load rather than by the individual rates.

To quantify traffic imbalance between the two users,
we introduce the asymmetry parameter$\alpha\;=\;\frac{\lambda_{A}}{\lambda_{A}+\lambda_{C}}\;\in\;[0,1]$.
The symmetric case corresponds to $\alpha=1/2$, while $\alpha\to1$
(or$\alpha\to0$) represents a highly asymmetric scenario in which
one user dominates the traffic. 

\subsection{Case 1:symmetric users at (approximately) critical
total load}

We begin with the symmetric scenario $\lambda_{A}=\lambda_{C}=4.5\times10^{4},\qquad\alpha=\frac{1}{2}$,
so the aggregate request rate is $\lambda_{{\rm tot}}=\lambda_{A}+\lambda_{C}=9.0\times10^{4}$.
For the baseline fixed operating point ($N_{{\rm tot}},\tau$) considered
here, this value places the system close to the critical boundary$\rho_{{\rm tot}}\approx1,\text{i.e}.\rho_{{\rm tot}}=(\lambda_{A}+\lambda_{C})\,\mathbb{E}[T(N_{{\rm tot}},\tau)]\approx1$.
\begin{figure*}
\begin{centering}
\begin{tabular}{|c|c|c|}
\hline 
\includegraphics[scale=0.17]{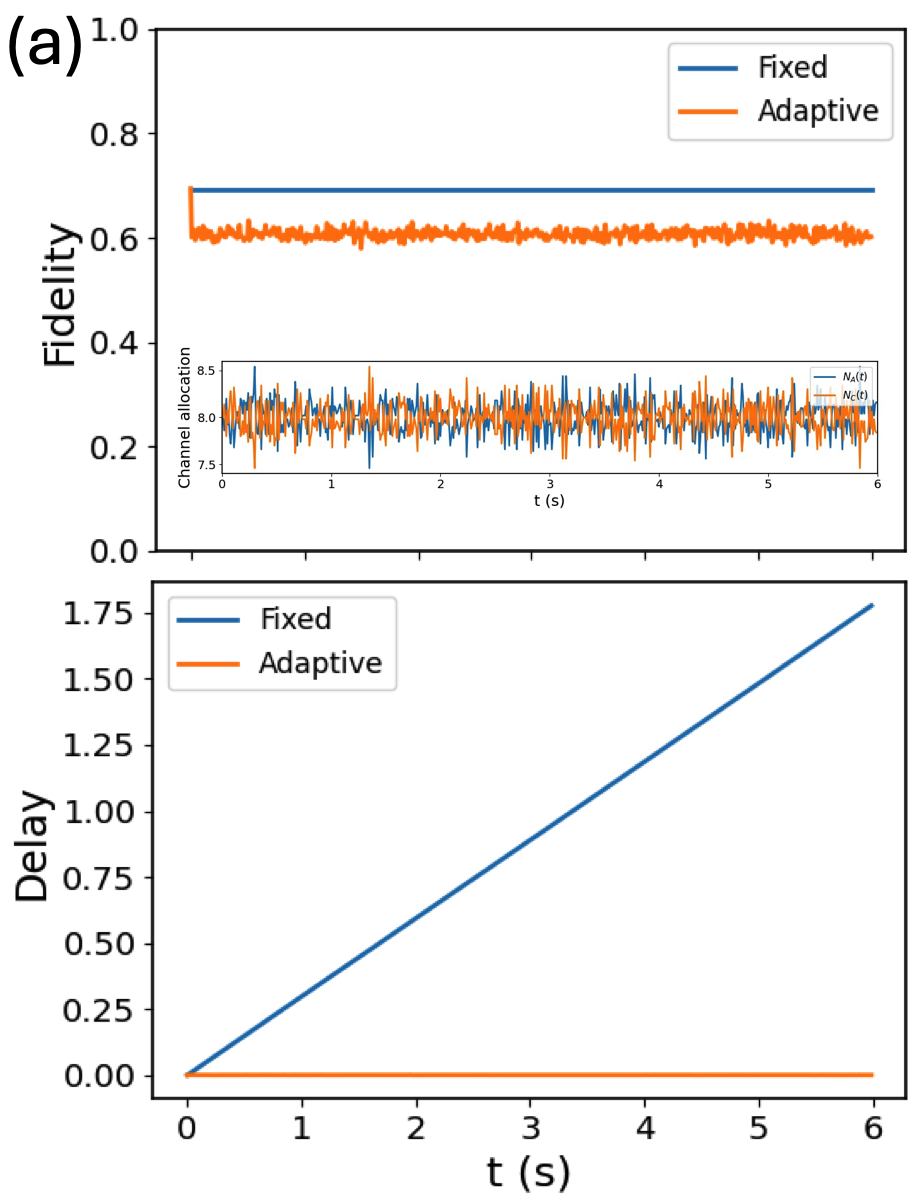} & \includegraphics[scale=0.155]{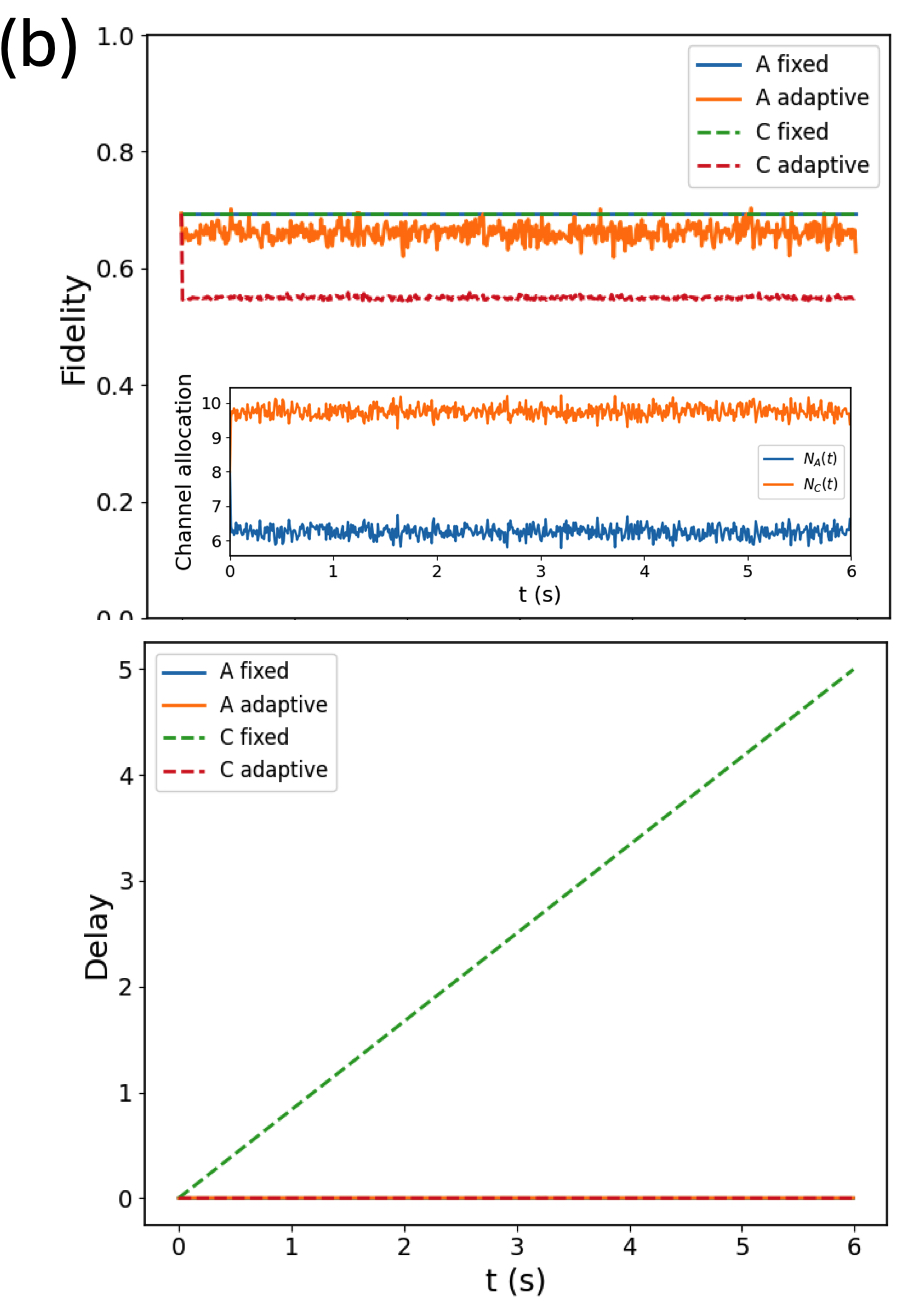} & \includegraphics[scale=0.166]{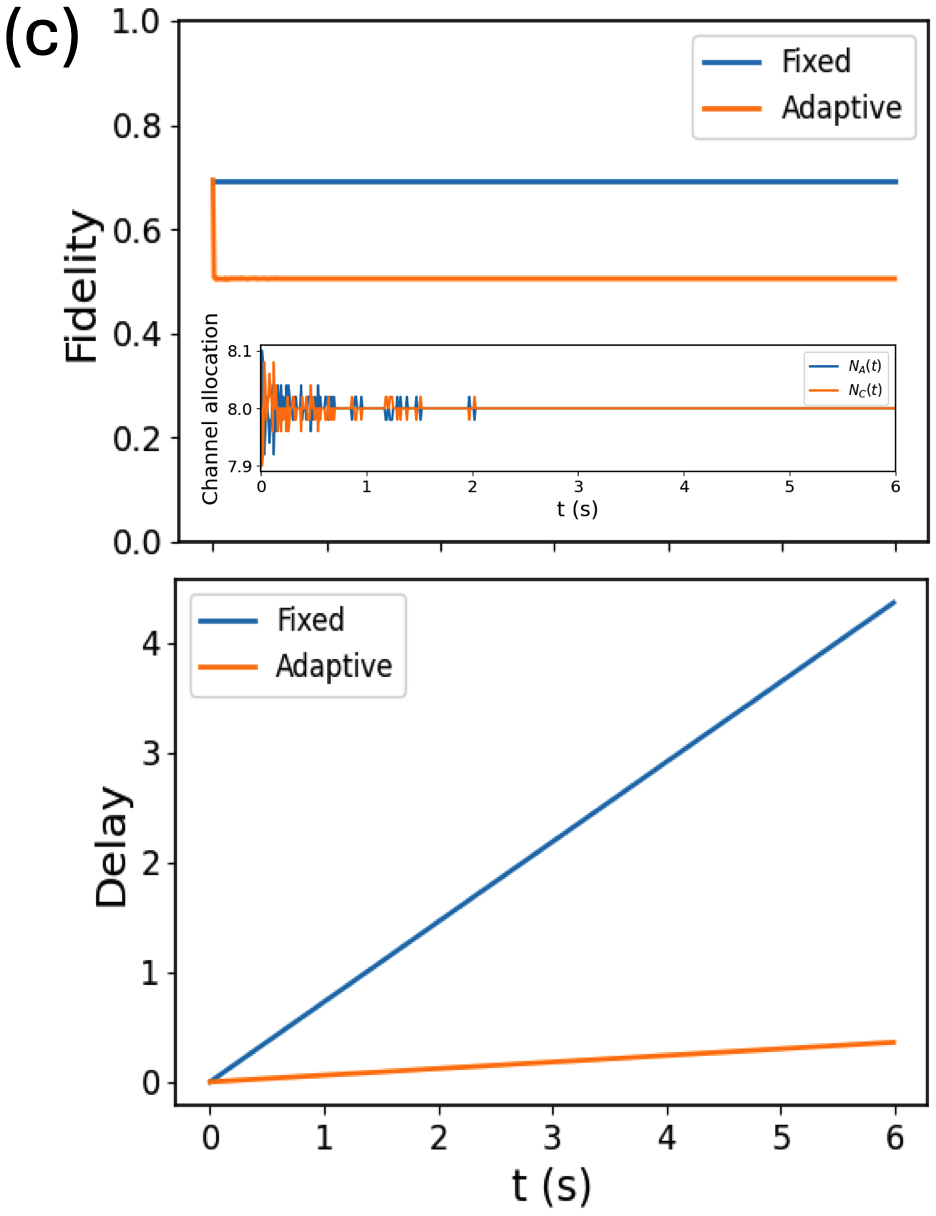}\tabularnewline
\hline 
\end{tabular}
\par\end{centering}
\caption{\label{fig:case1}Comparison between fixed and queue-aware adaptive
cutoff policies for two users under different traffic conditions.
Panels (a)--(c) show the temporal evolution of fidelity (left) and
delay (right). Insets (left) report the dynamic channel allocation
$N_{A}(t)$ and $N_{C}(t)$.(a) Symmetric traffic at critical load,
with $\lambda_{A}=\lambda_{C}=4.5\times10^{4}$ ($\rho_{\mathrm{tot}}\approx1$).(b)
Asymmetric traffic at critical load, with $\lambda_{A}=0.3\,\lambda_{\mathrm{tot}}$
and $\lambda_{C}=0.7\,\lambda_{\mathrm{tot}}$.(c) Symmetric traffic
in the supercritical regime, with $\lambda_{A}=\lambda_{C}=6\times10^{4}$
($\rho_{\mathrm{tot}}>1$).
}

\end{figure*}
Figure \ref{fig:case1} (a) reports the time evolution
of fidelity, delay, adaptive channel allocation, and cutoff time for
both policies. 

In the fixed-split case, each user receives a constant
allocation $N_{A}=N_{C}=N_{{\rm tot}}/2$ and a fixed cutoff $\tau$.
Because the system operates close to criticality, the effective service
capacity is only marginally sufficient, placing the system at the
stability boundary. As a result, the queues grow and the delay diverges,
although the fidelity remains bounded by the imposed cutoff.

In contrast, under the adaptive policy the cutoff
time increases dynamically in response to queue growth. This effectively
increases the service rate and shifts the operating point into the
subcritical region $\rho_{{\rm tot}}<1$, resulting in bounded queue
dynamics. As a consequence, both users exhibit a stable fidelity trajectory
that remains equal (by symmetry) and, for this operating point, exceeds
that of the fixed policy. The adaptive mechanism thus avoids the instability-induced
performance degradation observed in the fixed case. The difference
between the two policies is most clearly visible in the delay behavior. 

The adaptive channel allocation $N_{A}(t),N_{C}(t)$
exhibits small oscillations around the symmetric point $N_{A}(t)\approx N_{C}(t)\approx\frac{N_{{\rm tot}}}{2}$.
Although the instantaneous values fluctuate due to stochastic arrivals,
the time average converges to approximately 8 channels per user (for
$N_{{\rm tot}}=16$). This confirms that, in the symmetric traffic
case, the adaptive mechanism does not introduce long-term bias: resources
are shared evenly in expectation. The observed fluctuations are therefore
not structural asymmetries but the natural response to stochastic
queue imbalance.

The time evolution of the adaptive cutoff $\tau(t)$
reveals the stabilization mechanism explicitly. As the queues increase,
$\tau(t)$ is raised above its nominal value, effectively increasing
the mean service rate. Once the queues are stabilized, $\tau(t)$
remains at a level that balances stability and fidelity preservation.
This confirms that the load-regulation mechanism identified in the
single-user case extends to the symmetric multi-user setting.

\subsection{Case 2: Asymmetric users at fixed total load}

We now turn to the asymmetric traffic scenario,
in which the two users inject different arrival rates while sharing
the same repeater resources. Specifically, we fix the total arrival
rate at the critical value identified in the symmetric case, $\lambda_{{\rm tot}}=\lambda_{A}+\lambda_{C}\simeq9.1\times10^{4}$,
and introduce an imbalance between the users by choosing $\lambda_{A}=0.3\,\lambda_{{\rm tot}},\qquad\lambda_{C}=0.7\,\lambda_{{\rm tot}}$,
corresponding to an asymmetry parameter $\alpha=0.3.$

In contrast to the symmetric case, the two queues
are now driven by different stochastic inputs, which leads to distinct
fidelity and delay dynamics for Alice and Charlie. Figure \ref{fig:case1}
(b) shows the time evolution of fidelity, delay, channel allocation,
and cutoff times for the asymmetric configuration. Under the fixed
split policy, both users receive a constant allocation $N_{A}=N_{C}=N_{{\rm tot}}/2$.
Because Charlie carries 70\% of the total load, this rigid partition
is insufficient to serve his queue at the critical aggregate load.
As a result, the Charlie queue diverges, leading to unstable delay
behavior. Alice, by contrast, is lightly loaded and remains stable.
Under the adaptive policy, resources are dynamically reallocated.
The controller assigns on average fewer channels to Alice (approximately
6) and correspondingly more to Charlie. This redistribution shifts
Charlie\textquoteright s effective load below unity, resulting in
bounded queue growth over time. The impact of this reallocation is
directly visible in the fidelity traces. Because resources are reallocated
toward the heavily loaded user, Alice operates with fewer channels
than under fixed partitioning, leading to a modest fidelity reduction.
Charlie receives more channels, but must also operate with larger
cutoff times to maintain stability. As a result, fidelity becomes
user-dependent under adaptation, reflecting congestion-driven redistribution
of service capacity.

The delay curves further highlight the role of dynamic
control. In the fixed split case, Charlie\textquoteright s delay grows
without bound due to insufficient allocated resources, while Alice\textquoteright s
delay remains stable. In contrast, adaptive reallocation transfers
capacity from the lightly loaded user to the heavily loaded one, restoring
bounded queue dynamics for both users. Charlie\textquoteright s delay
stabilizes instead of diverging, whereas Alice experiences only a
moderate increase relative to the symmetric case. Operation at the
critical aggregate load leads to divergence under rigid partitioning
but can be shifted into the subcritical regime through queue-aware
resource redistribution.

The channel allocation fluctuates around mean values
consistent with the traffic imbalance (\ensuremath{\approx}6 and \ensuremath{\approx}10
channels), confirming demand-proportional resource redistribution.

Overall, in the asymmetric near-critical regime,
adaptive control acts as a congestion-driven redistribution mechanism:
channels are withdrawn from the lightly loaded user and reassigned
to the heavily loaded one. Stability is thereby restored for the dominant
flow, at the expense of moderate performance degradation for the lighter
flow. The fixed policy, by contrast, cannot react to traffic imbalance
and therefore fails to stabilize the heavily loaded queue, even though
the total offered load remains unchanged. This illustrates that stability
in shared systems depends not only on total load but also on its distribution
across flows.

The observed resource redistribution can be interpreted
more broadly in terms of fairness criteria in shared queueing systems.
A fixed partition enforces equal resource allocation across users,
independent of their traffic demand, corresponding to an equal-rate
or equal-resource notion of fairness. By contrast, the adaptive policy
implements a congestion-responsive allocation in which resources are
dynamically reassigned based on queue pressure.

In this sense, the controller does not aim to equalize
throughput or fidelity across users, but rather to regulate queue
growth by prioritizing the more heavily loaded flow. This results
in a form of demand-proportional or congestion-aware fairness, in
which service capacity is allocated to maintain bounded delay whenever
the aggregate load permits. The lightly loaded user experiences a
modest degradation in performance, while the heavily loaded user is
stabilized.

These results highlight that adaptive control implicitly
selects a fairness criterion: stability and bounded delay are prioritized
over equal resource allocation or equal performance across users.
This tradeoff is a fundamental consequence of operating near the stability
boundary: when total capacity is nearly saturated, it is not possible
to simultaneously maintain equal resource allocation and bounded delay
for users with asymmetric demand.

\subsection{Case 3: Symmetric users in the supercritical regime}

We finally consider a symmetric but clearly overloaded
configuration, $\lambda_{A}=\lambda_{C}=6.0\times10^{4}$, so that
the aggregate arrival rate significantly exceeds the effective service
capacity for the chosen ($N_{{\rm tot}},\tau$). In this case, the
system operates in the supercritical regime $\rho_{{\rm tot}}>1.$

Figure \ref{fig:case1} (c) shows the time evolution
of fidelity, delay, adaptive channel allocation and adaptive cutoff.

Because the total offered load exceeds the maximum
achievable service rate, neither the fixed nor the adaptive policy
can restore stability. The delay therefore diverges in both cases.
However, the divergence rate differs substantially. Under the fixed
policy, the queues grow rapidly, leading to a steep increase in delay.
Under adaptive control, the cutoff time increases in response to congestion,
effectively increasing the mean service rate and slowing the growth
of the queues. Although instability cannot be eliminated, the adaptive
mechanism significantly mitigates delay growth relative to static
operation.

Under symmetric overload, redistribution cannot
alter the aggregate capacity constraint. The controller therefore
converges to equal allocation, and cutoff adaptation becomes the only
available control dimension. Instability persists, but delay growth
is attenuated relative to fixed operation at the expense of reduced
fidelity.

\section{Conclusion and discussion}

In this work we investigated the interplay between
quantum-memory decoherence, stochastic traffic, and queue dynamics
in a minimal quantum repeater architecture. By combining a physically
grounded model of probabilistic entanglement generation and memory
decay with a queueing-theoretic description of entanglement requests,
we characterized how memory cutoff times and parallel generation resources
jointly determine fidelity, delay, and stability under both stationary
and time-varying traffic conditions.

Under fixed policies, system behavior is governed
by the load parameter $\rho=\lambda E[T].$ The cutoff time $\tau$
regulates a fundamental fidelity--latency trade-off, while the number
of channels N determines effective service capacity and therefore
stability. Near critical load, inappropriate parameter choices lead
to marginal drift or divergence of the queue, whereas in the supercritical
regime no fixed configuration can prevent congestion growth.

Adaptive control fundamentally changes this picture.
Cutoff adaptation stabilizes near-critical systems by increasing the
effective service rate during congestion, but does so at the cost
of reduced fidelity due to longer memory dwell times. Resource scaling,
by contrast, directly increases service capacity and can restore stability
in the critical regime without degrading entanglement quality, at
the expense of activating additional physical channels. Under bursty
ON--OFF traffic, adaptive strategies significantly attenuate delay
spikes relative to static operation. When both the cutoff time and
the number of channels are allowed to vary, congestion can be mitigated
while limiting fidelity degradation. Notably, joint adaptation does
not rely on permanent over-provisioning: in the supercritical burst
scenario considered, the time-averaged number of active channels was
approximately 12, lower than the fixed baseline of 16. Additional
resources are therefore activated only during congestion intervals
and released during low-traffic periods, indicating that queue-aware
control can simultaneously improve temporal performance and reduce
average hardware utilization.

Extending the framework to a minimal two-user scenario
in which independent traffic flows share a common pool of entanglement-generation
resources reveals an additional structural insight. Stability is governed
by the total offered load $\rho_{\mathrm{tot}}=(\lambda_{A}+\lambda_{C})E[T]$,
but performance becomes user-dependent when traffic is asymmetric.
Under rigid resource partitioning, a heavily loaded user may experience
queue divergence even when the aggregate load lies at the critical
boundary. Adaptive control dynamically redistributes channels according
to congestion levels, transferring capacity from lightly loaded users
to heavily loaded ones and restoring bounded delay. This stabilization
mechanism induces user-dependent fidelity and delay behavior, reflecting
the congestion-driven redistribution of service capacity. In multi-user
settings, fairness should therefore be understood not as equal resource
partitioning, but as balanced service performance under asymmetric
demand.

Taken together, these results show that quantum
repeater nodes operating under realistic traffic must be viewed as
congestion-sensitive service systems in which physical-layer parameters
directly influence queue stability. Memory decoherence couples fidelity
to temporal performance in a manner absent from classical networks,
and resource allocation decisions determine not only throughput but
also entanglement quality. Even simple, local queue-aware feedback
laws provide substantial robustness gains without requiring traffic
prediction or global coordination. This suggests that lightweight
adaptive control mechanisms may already be sufficient to enhance the
operational stability and hardware efficiency of early quantum network
deployments. From a system-design perspective, these results suggest
that quantum repeater nodes should not rely solely on fixed protocol
parameters, but instead benefit from queue-aware adaptation of memory
cutoff times and generation resources. In particular, cutoff adaptation
provides a low-overhead mechanism for stabilizing operation near critical
load, while dynamic resource scaling is required to maintain both
stability and fidelity under bursty or multi-user demand.

Future extensions include multi-hop repeater chains,
heterogeneous links and memories, finite buffer constraints, and
coordinated multi-node control policies. In larger network topologies,
the interaction between cutoff control, resource scaling, and routing
decisions is likely to play a central role in determining end-to-end
performance. Overall, this work establishes queue-aware
control as a fundamental component of quantum network operation under
stochastic demand.


\section*{Acknowledgements}

The authors thank Dr. Hon Wai Lau for fruitful discussions. This work
was supported by the Moonshot R\&D Program Grant JPMJMS226C and the
JSPS KAKENHI Grants No. JP21H04880 and No. JP24K07485.

\bibliographystyle{ieeetr}
\bibliography{bib1}

\appendix

\section*{\label{sec:Derivation-of-the}Derivation of the mean service time
$\mathbb{E}[T(N,\tau)]$}

The service time $T(N,\tau)$ used in the queueing model is derived
from the underlying two-link entanglement-generation process with
cutoff. Each elementary link is generated according to a Poisson process
with effective rate $\mu(N)=NRp_{\mathrm{gen}}$, where $N$ is the
number of channels used, $R$ is the repetition rate, and $p_{\mathrm{gen}}$
is the success probability per channel. Service completion requires
two elementary links. The first link is obtained after a waiting time
$\min(X_{1},X_{2})$, where $X_{1},X_{2}\sim\mathrm{Exp}(\mu(N))$,
giving mean $1/[2\mu(N)]$. Once the first link is stored, the system
waits for the second link for at most a cutoff time $\tau$. If the
second link arrives within $\tau$, service is completed; otherwise
the stored link is discarded and the process restarts. Writing $Z\sim\mathrm{Exp}(\mu(N))$
for the waiting time to the second link and $q=e^{-\mu(N)\tau}$ for
the cutoff-failure probability, the mean service time satisfies the
renewal equation:
\begin{equation}
\mathbb{E}[T(N,\tau)]=\frac{1}{2\mu(N)}+\mathbb{E}[\min(Z,\tau)]+q\,\mathbb{E}[T(N,\tau)].
\label{eq:mean_service_time}
\end{equation}
Since $\mathbb{E}[\min(Z,\tau)]=\int_{0}^{\tau}\Pr(Z>t)\,dt=\int_{0}^{\tau}e^{-\mu(N)t}\,dt=\frac{1-e^{-\mu(N)\tau}}{\mu(N)}$,
this yields $\mathbb{E}[T(N,\tau)]=\frac{\frac{3}{2}-e^{-\mu(N)\tau}}{\mu(N)\bigl(1-e^{-\mu(N)\tau}\bigr)}$.
The effective service rate is therefore $S(N,\tau)=1/\mathbb{E}[T(N,\tau)]$,
which is the quantity entering the queueing dynamics and the stability
condition $\lambda\mathbb{E}[T(N,\tau)]<1$.

To evaluate $\mathbb{E}[\min(Z,\tau)]$, we use the identity $\mathbb{E}[X]=\int_{0}^{\infty}\Pr(X>t)\,dt$,
valid for any non-negative random variable $X$. So if we apply it
to $X=\min(Z,\tau)$, then $\mathbb{E}[\min(Z,\tau)]=\int_{0}^{\infty}\Pr(\min(Z,\tau)>t)\,dt$.

To evaluate $\Pr(\min(Z,\tau)>t)$ we can have two cases: if $t<\tau$
then $\Pr(\min(Z,\tau)>t)=\Pr(Z>t)$. If $t>\tau$, then $\min(Z,\tau)\le\tau<t$,
so $\Pr(\min(Z,\tau)>t)=0$.
\EOD
\end{document}